\documentclass[useAMS,usenatbib]{mn2e}
\usepackage{amsmath}
\usepackage{amssymb,amsfonts}
\usepackage{graphicx,epsfig,graphics}
\usepackage{graphics}
\usepackage{ulem}
\newcommand{\bt}{\mbox{\boldmath{$\theta$}}}

\newcommand{\bx}{{\boldsymbol{\vartheta}}}

\newcommand{\map}{M_{\rm ap}}

\begin{document}

\label{firstpage}
\title[Aperture Mass Statistic]{Fast Calculation of the  Weak Lensing Aperture Mass Statistic} 
\author[A. Leonard, S. Pires, J.-L. Starck] {Adrienne
  Leonard\thanks{Email: adrienne.leonard@cea.fr}, Sandrine Pires, and Jean-Luc Starck \\ Laboratoire AIM, UMR CEA-CNRS-Paris 7, Irfu, SAp/SEDI, Service d'Astrophysique, \\CEA Saclay, F-91191 GIF-SUR-YVETTE CEDEX, France.}

\date{}
\maketitle

\pagerange{\pageref{firstpage}--\pageref{lastpage}} \pubyear{2009}

\begin{abstract}
{The aperture mass statistic is a common tool used in weak lensing studies. By convolving lensing maps with a filter function of a specific scale, chosen to be larger than the scale on which the noise is dominant, the lensing signal may be boosted with respect to the noise. This allows for detection of structures at increased fidelity. Furthermore, higher-order statistics of the aperture mass (such as its skewness or kurtosis), or counting of the peaks seen in the resulting aperture mass maps, provide a convenient and effective method to constrain the cosmological parameters. In this paper, we more fully explore the formalism underlying the aperture mass statistic. We demonstrate that the aperture mass statistic is formally identical to a wavelet transform at a specific scale. Further, we show that the filter functions most frequently used in aperture mass studies are not ideal, being non-local in both real and Fourier space. In contrast, the wavelet formalism offers a number of wavelet functions that are localized both in real and Fourier space, yet similar to the ÔoptimalÕ aperture mass filters commonly adopted. Additionally, for a number of wavelet functions, such as the starlet wavelet, very fast algorithms exist to compute the wavelet transform. This offers significant advantages over the usual aperture mass algorithm when it comes to image processing time, demonstrating speed-up factors of $\sim 5 - 1200$ for aperture radii in the range $2$ to $64$ pixels on an image of $1024 \times 1024$ pixels.}
\end{abstract}


\section{Introduction}

The aperture mass statistic, $\map$, is a well-established tool in weak gravitational lensing studies, being useful for detection of structures \citep{S96, Schneideretal98} and for the determination of cosmological parameters through related statistics such as its variance, skewness and kurtosis, or through peak statistics \citep[e.g.][]{Schneideretal98, Jarvisetal04, KS05, DH10}. Formally, the aperture mass technique consists of applying a filter, defined on circular apertures, to a map of measured shears. 

The strength of the aperture mass technique lies in the fact that the structures responsible for the lensing signal are dominant over the noise at certain scales. Choosing an aperture filter whose shape traces that of the expected signal and with a scale similar to that of the expected structures results in an aperture mass map with optimal signal to noise. 
Moreover, assessment of the noise properties of a given $\map$ reconstruction is straightforward, carried out by randomising the input data a large number of times and computing the variance of the resulting $\map$ reconstructions, $\sigma^2_{M_{ap}}$. The signal-to-noise of the aperture mass map can thereby be directly computed from the data : $S = \frac{M_{ap}}{\sigma_{M_{ap}}}$.

However, the signal-to-noise will only be maximum in the aperture mass map if the filter function is tuned to follow the shape of the expected signal, which is not necessarily known \textit{a priori}, and when the aperture size matches the scale of the structures we aim to detect. Choosing an ideal filter function is, consequently, somewhat challenging. Perhaps the largest difficulty with this method, though, is that evaluating the $\map$ statistic and its associated noise map is time consuming, particularly for larger apertures, and this limits the ability to consider reconstructions of large fields, or across a range of scales.

An alternative method for analysing weak lensing data at multiple different scales simultaneously is to use the wavelet formalism \citep{SPR06}. 
This method involves first converting the shear map, evaluated via measurement of galaxy shapes, into a convergence map in Fourier space, then applying a wavelet transform. One advantage of this method is that there exist fast algorithms to compute the wavelet transform. These algorithms compute simultaneously several wavelet bands analogous to aperture mass maps at dyadic scales.

In this paper, we demonstrate that the aperture mass statistic is formally equivalent to the wavelet transform of a map of the weak lensing convergence $\kappa$ evaluated at a particular scale, for appropriate choice of filter function. Furthermore, we demonstrate that the most frequently used aperture mass filter functions \citep{Schneideretal98, vanWaerbeke98,Jarvisetal04} do not meet all of the desired requirements of aperture mass filters, and are therefore not ideal. We propose an alternative filter function -- the starlet wavelet -- and demonstrate that this filter function meets all the desired requirements for aperture mass filter functions. Moreover, we show that this function is very similar in form to the aperture mass filter function found to be optimal for weak lensing skewness studies in \citet{Zhang03}, and is therefore a perfect choice for lensing studies.  Lastly, we demonstrate that the {\it \`{a} trous} wavelet transform algorithm, publicly available as part of the MRLens software package\footnote{http://irfu.cea.fr/Ast/mrlens\_software.php} \citep{SPR06}, is substantially faster than the aperture mass algorithm, offering up to three orders of magnitude speed-up in processing time for an $1024\times1024$ pixel image. 

This paper is organised as follows. In \S~\ref{sec:map}, we describe the formalism underlying the aperture mass method, and outline the desired characteristics of the associated filter functions. In \S~\ref{sec:wavelet}, we define the wavelet transform of an image, and demonstrate that the wavelet transform of a map of the weak lensing convergence $\kappa$ is identical to the aperture mass statistic evaluated at corresponding scales. In \S~\ref{sec:filters}, we describe the two most commonly-used aperture mass filter functions, and compare these to the starlet wavelet function, which we show to be more ideally-suited for aperture mass studies. In \S~\ref{sec:timing}, we compare the processing time for the aperture mass and wavelet transform algorithms, and we conclude with a summary of our results in \S~\ref{sec:summary}.

\section{The aperture mass statistic}
\label{sec:map}

\subsection{$\map$ definition}
The aperture mass statistic is defined \citep{S96} as the convolution of the convergence, $\kappa$, with a radially symmetrical filter function, $U(|\bx|)$:

\begin{equation}
\map(\bt) = \int d^2\boldsymbol{\vartheta}\ \kappa(\boldsymbol{\vartheta}) U(|\boldsymbol{\vartheta}|)\ .
\label{eq:mapdef}
\end{equation}
By considering the relationship between the shear, $\gamma$, and the convergence, one can reformulate equation \eqref{eq:mapdef} in terms of the measured shear as
\begin{equation}
\map(\bt) = \int d^2\boldsymbol{\vartheta}\ \gamma_t(\bx) Q(|\bx|)\ ,
\label{eq:shearmap}
\end{equation}
where $\gamma_t(\bx)$ is the tangential component of the shear at position $\bx$ relative to the centre of the aperture, $Q(|\bx|)$ is a second radially-symmetric function, related to $U(|\bx|)$ by:
\begin{equation}
Q(\vartheta) \equiv \frac{2}{\vartheta^2}\int_0^\vartheta \vartheta^\prime U(\vartheta^\prime) d\vartheta^\prime - U(\vartheta)\ ,
\label{eq:QU}
\end{equation}
and $U(\vartheta)$ is subject to the condition that 
\begin{equation}
\int_0^\theta \vartheta\ U(\vartheta) \ d\vartheta = 0\ ,
\label{eq:comp}
\end{equation}
with $\theta$ being the radius of the aperture. Furthermore, $Q(\vartheta)$ and $U(\vartheta)$ are required to go to zero smoothly at $\vartheta=\theta$. It is also preferable that the power spectrum of $U(\vartheta)$ is local in the frequency domain, and shows no oscillatory behaviour. This ensures that the filter function acts as a band-pass filter, allowing detection of structures at the scale of interest only.

The filter functions $Q(\vartheta)$ and $U(\vartheta)$ may take any form, subject to the conditions outlined above, namely:
\begin{enumerate}
\item The filter function should be compensated within the aperture being considered.
\item The filter function should be local in real space; i.e. it should go to zero smoothly at a finite radius, and be zero outside this radius. 
\item The filter function should be local in Fourier space, with no oscillatory behaviour in the power spectrum.
\end{enumerate}

\citet{S96} showed that optimal signal to noise will be achieved if the function $Q(\vartheta)$ matches the expected profile of $\gamma_t$  as far as possible. This is generally not known \textit{a priori}, so a number of generic functions, or families of functions, have been proposed \citep[see, e.g.][]{Schneideretal98, vanWaerbeke98, Jarvisetal04}. These popular filter functions will be discussed in detail in \S~\ref{sec:filters} below, where we will demonstrate that none of them are able to simultaneously meet all three of the above criteria. 

We now consider the wavelet transform and its application in weak lensing. Specifically, we will demonstrate below that the wavelet transform at a given scale is formally equivalent to the aperture mass statistic described above.

\subsection{Associated Statistics of $\map$}

The aperture mass statistic is useful not only for generating filtered maps of the lensing convergence, and thereby detecting structures as a function of aperture scale, but also for constraining cosmological parameters through associated statistics of the aperture mass $\map$. 

Most commonly, the variance of the aperture mass statistic, $\left\langle \map^2\right\rangle(\theta)$, is considered as a function of aperture radius $\theta$. This measure is related to the power spectrum $P_\kappa(k)$ via \citep{Schneideretal98}:
\begin{equation}
\left\langle\map^2\right\rangle(\theta) = \frac{1}{2\pi}\int d^2\boldsymbol{k}P(k)W(k\theta)\ ,
\label{eq:varmap}
\end{equation}
where
\begin{equation}
W(k\theta) = [\tilde{U}(k)]^2
\end{equation}
is the power spectrum of the filter $U(\vartheta)$.

Equation \eqref{eq:varmap} can equivalently be expressed in terms of the two point correlation function of the shear \citep[see, e.g.][]{Crittenden02, Schneideretal02, Jarvisetal04}, and this statistic helps to constrain the amplitude of the matter power spectrum $\sigma_8$.

In order to probe non-gaussianities in our cosmological model, higher-order statistics need to be considered. \cite{Jarvisetal04} present a derivation of the skewness of the aperture mass statistic $\left\langle\map^3\right\rangle$ in terms of the Bispectrum or, equivalently, the three-point correlation function of the shear. This statistic, as well as the kurtosis of the convergence and aperture mass, and peak counting are considered in \cite{Pires09a, pires11}, where the efficacy of these probes to discriminate between competing cosmological models along the $\Omega_M-\sigma_8$ degeneracy are considered.

\section{Wavelet Transform}
\label{sec:wavelet}

The wavelet transform is a multiscale transform, where the wavelet coefficients of an image are computed at each position in the image at various different scales simultaneously. In one dimension, the wavelet coefficient of a function $f(x)$, evaluated at position $b$ and scale $a$ is defined as \citep{smb98, starck:book06}:
\begin{equation}
W(a,b) = \frac{1}{\sqrt{a}}\int\  f(x)\psi^\ast\left(\frac{x-b}{a}\right)\ dx\ ,
\label{eq:defwl}
\end{equation}
where $\psi(x)$ is the analysing wavelet. The analysis is analogous in 2 dimensions, with $\psi(x,y) = \psi(x)\psi(y)$. 

By definition, wavelets are compensated functions; i.e. the wavelet function $\psi(x)$ is constrained such that$\int_{\mathbb{R}^1} \psi(x)dx = 0$ and hence, by extension 
\begin{equation}
\iint_{\mathbb{R}^2}\psi(x,y) dx\ dy = 0.
\end{equation}
According to the definition in equation \eqref{eq:defwl}, the continuous wavelet transform of an image is therefore nothing more than the convolution of that image with compensated filter functions of various characteristic scales. If the image $f(x,y)$ is taken to be the convergence $\kappa(x,y)$, then for an appropriate choice of (radially-symmetric, local) wavelet, the wavelet transform is formally identical to the aperture mass statistic at the corresponding scales.

In practice in application, we use the starlet transform algorithm \citep[see][]{smb98, starck:book06}, which simultaneously computes the wavelet transform on dyadic scales corresponding to $2^j$ pixels. This algorithm decomposes the convergence map of size $N \times N$ into $J = j_{\max}+1$ sub-arrays of size $N\times N$ as follows:
\begin{equation} 
\kappa(x,y) = c_{J}(x,y) + \sum_{j=1}^{j_{\max}}w_j(x,y)\ ,
\end{equation}
where $j_{\max}$ represents the number of wavelet bands (or, equivalently, aperture mass maps) considered, $c_{J}$ represents a smooth (or continuum) version of the original image $\kappa$, and $w_j$ represents the input map filtered at scale $2^j$ (i.e. the aperture mass map at $\theta = 2^j$ pixels).

Using the wavelet formalism to derive the aperture mass statistic presents different advantages:
\begin{itemize}
\item Many families of wavelet functions have been studied in the statistical literature, and all these wavelet functions could be applied to the aperture mass statistic. We will demonstrate in \S~\ref{sec:filters} that the starlet wavelet, in particular, simultaneously meets all three of the requirements for aperture mass filters described in \S~\ref{sec:map}, and is therefore ideal for weak lensing studies.
\item For some specific wavelet functions, discrete and very fast algorithms exist, allowing us to compute a set of wavelet scales through the use of a filter bank with a very 
limited number of operations. See \citet{starck:book10} for a full review of the different wavelet transform algorithms.
\item Wavelets have shown to be very powerful for non Gaussianity studies  for both weak lensing data \citep{SPR06,Pires09a,pires11} and CMB data 
\citep{gauss:vielva04,starck:jin05,starck:sta03_1,vielva06, McEwen08a,martinez11}.
\end{itemize}

\section{Filter Functions}
\label{sec:filters}

We have shown above that a 2D wavelet transform of a map of the convergence $\kappa$ is formally identical to computing aperture mass maps of the convergence -- or, equivalently, the shear -- at the chosen scales. We now consider the functional form of the compensated filter functions commonly associated with aperture mass studies, and compare with a wavelet transform that has been frequently used for weak lensing studies: the starlet wavelet \citep{SPR06, Pires09a,pires11}.

\subsection{Aperture Mass Filter Functions}
\cite{Schneideretal98} have proposed a family of polynomial filter functions defined by:
\begin{eqnarray}
U(\vartheta) &=& \frac{(\ell+2)^2}{\pi\theta^2} \left(1-\frac{\vartheta^2}{\theta^2}\right)^\ell\left(\frac{1}{\ell+2}-\frac{\vartheta^2}{\theta^2}\right)\ \mathcal{H}(\theta-\vartheta),\nonumber\\
Q(\vartheta) &=& \frac{(1+\ell)(2+\ell)}{\pi\theta^2}\frac{\vartheta^2}{\theta^2}\left(1-\frac{\vartheta^2}{\theta^2}\right)^\ell\mathcal{H}(\theta-\vartheta)\ ,
\label{eq:sch}
\end{eqnarray}
where $\mathcal{H}(\theta-\vartheta)$ is the Heaviside step function, which takes the value of 1 for $\vartheta\le\theta$ and zero for $\vartheta>\theta$. 

More recently, several authors \citep[e.g.][]{vanWaerbeke98, Jarvisetal04} have advocated a filter function of the form:
\begin{eqnarray}
U(\vartheta) &=& \frac{A}{\theta^2}\left(1-\frac{b\vartheta^2}{\theta^2}\right)\exp\left(-\frac{b\vartheta^2}{\theta^2}\right)\ ,\nonumber\\
Q(\vartheta) &=& \frac{A}{\theta^2} \frac{b\vartheta^2}{\theta^2}\exp\left(-\frac{b\vartheta^2}{\theta^2}\right)\ ,
\label{eq:mexhat}
\end{eqnarray}
where various choices for the constant $b$ and the overall normalisation $A$ have been used in the literature \citep[see, e.g.,][]{Crittenden02, Zhang03, Jarvisetal04}. Of specific interest to this paper are the forms used by \citet{vanWaerbeke98}, who takes $b=4$, and that of \citet{Jarvisetal04} and \citet{Zhang03}, who take $b=1/2$.\footnote{This form is analogous to the Mexican Hat wavelet function} 

\citet{Zhang03} have shown that the form in equation \eqref{eq:mexhat} is optimal for weak lensing skewness measurements, and \cite{vanWaerbeke98} notes that this form is preferable to the polynomial filter of \cite{Schneideretal98} for skewness measurements because the power spectrum of $U(\vartheta)$ is non-oscillatory, with a well-localised peak in the frequency domain. 

Note that the $Q(\vartheta)$ filter function described in equation \eqref{eq:mexhat} shows a peak at $\vartheta = \sqrt{2}\theta$, and tends to zero as $\vartheta \rightarrow \infty$. In practice, any algorithm used to generate an aperture mass map will need to truncate this filter function at some finite radius. This will involve a trade-off between accuracy and algorithm speed, and truncation may affect the effective width of the filter. This will be discussed more fully in sections \ref{subsec:real}, \ref{subsec:fourier}, and \ref{sec:timing} below.

\subsection{Starlet Wavelet Function}
One of the most popular wavelet transform algorithms in astronomy is the Isotropic Undecimated Wavelet Tranform (IUWT) \citep{starck:book10, starck:book11}, also called {\it {\`a} trous} algorithm or starlet transform. 
In this algorithm, the  wavelet $\psi(x,y)$ is separable and can be defined by:
\begin{equation}
\psi\left(\frac{x}{2},\frac{y}{2}\right) = 4\left[\varphi(x,y) - \frac{1}{4}\varphi\left(\frac{x}{2},\frac{y}{2}\right)\right]\ ,
\label{eq:splinewl}
\end{equation}
where $\varphi(x,y) = \varphi(x)\varphi(y)$ and  $\varphi(x)$ is a scaling function from which the wavelet is generated. In the case of the starlet wavelet, $\varphi(x)$ is a B3-spline:
\begin{equation}
\varphi(x)=\frac{1}{12}(|x-2|^3-4|x-1|^3+6|x|^3-4|x+1|^3+|x+2|^3),
\end{equation}
which is a compact function that is identically zero for $|x|>2$. 
  
This wavelet function has a compact support in real space, is well localized in Fourier domain, and the wavelet decomposition of an image can be obtained with a very fast algorithm (see \citet{starck:book06} for a full description). Computation time will be discussed in detail in Section~\ref{sec:timing}.

Evaluation of equation \eqref{eq:splinewl} yields the following aperture mass filters: 
\begin{eqnarray}
U(\eta) = \frac{1}{9}\left(93\left|\eta\right|^3-64\left[\left|\frac{1}{2}-\eta\right|^3+\left|\frac{1}{2}+\eta\right|^3\right]\right.\nonumber\\ 
 \left.+18\left[\left|1-\eta\right|^3+\left|1+\eta\right|^3\right]-\frac{1}{2}\left[\left|2-\eta\right|^3+\left|2+\eta\right|^3\right]\right)\ , 
\end{eqnarray}\normalsize
where $\eta = \vartheta/\theta$, and
\begin{eqnarray}
Q(\eta) = \frac{1}{45\eta^2}\left(-279\eta^2|\eta|^3+8[1+6\eta+24\eta^2]\left|\frac{1}{2}-\eta\right|^3\right. \nonumber\\ \left.-9[1+3\eta+6\eta^2]\left|1-\eta\right|^3-9[1-3\eta+6\eta^2]\left|1+\eta\right|^3\right.\nonumber\\ \left.+\left[1+\frac{3}{2}\eta+\frac{3}{2}\eta^2\right]\left|2-\eta\right|^3+{\left[1-\frac{3}{2}\eta+\frac{3}{2}\eta^2\right]|2+\eta|^3} \right.\nonumber\\ \left.+ 8[1-6\eta+24\eta^2]\left|\frac{1}{2}+\eta\right|^3\right)\ .
\end{eqnarray}\normalsize

\subsection{Properties in Real Space}
\label{subsec:real}

\begin{figure}
\centering{
\includegraphics[width=0.5\textwidth]{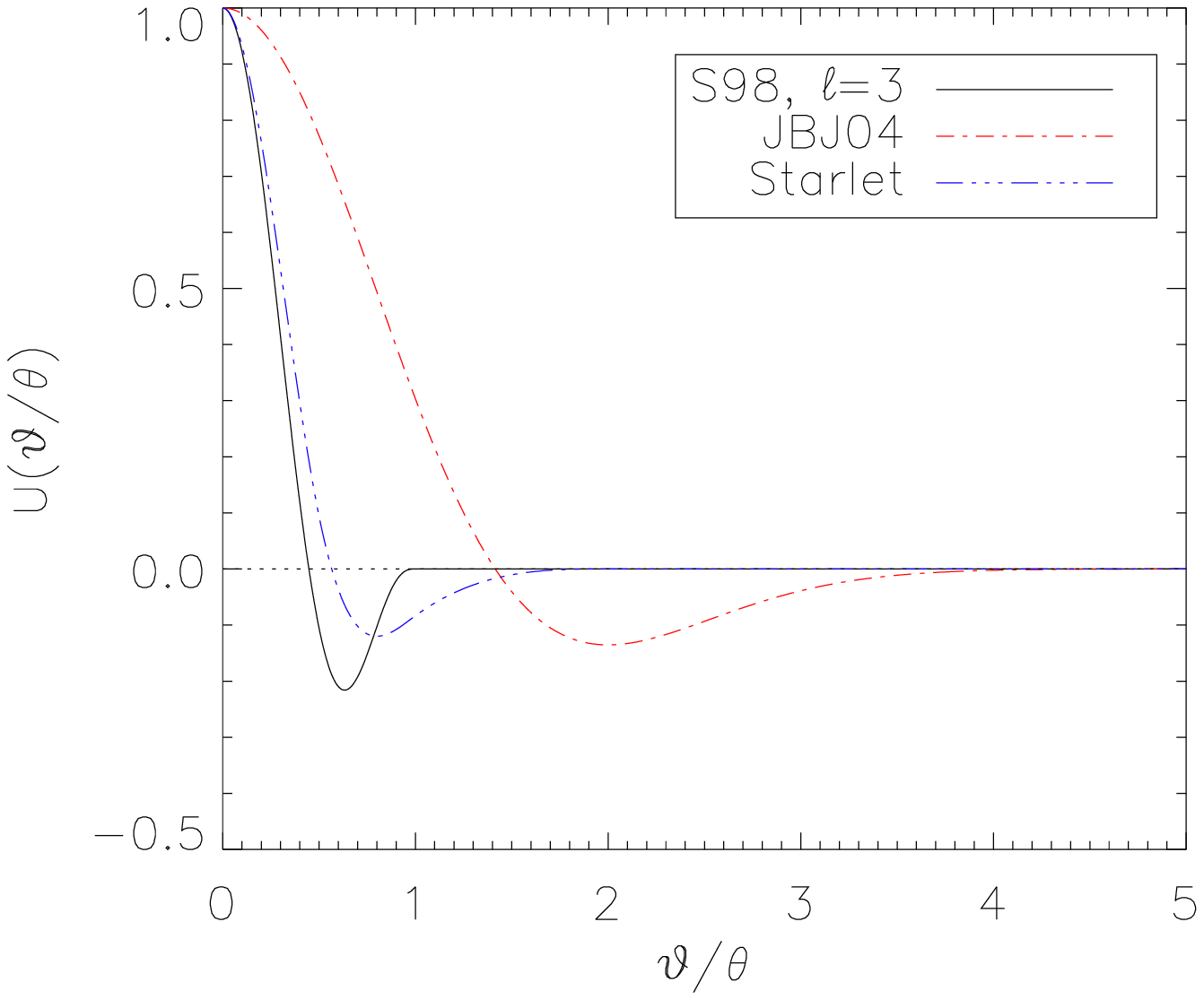}
\includegraphics[width=0.5\textwidth]{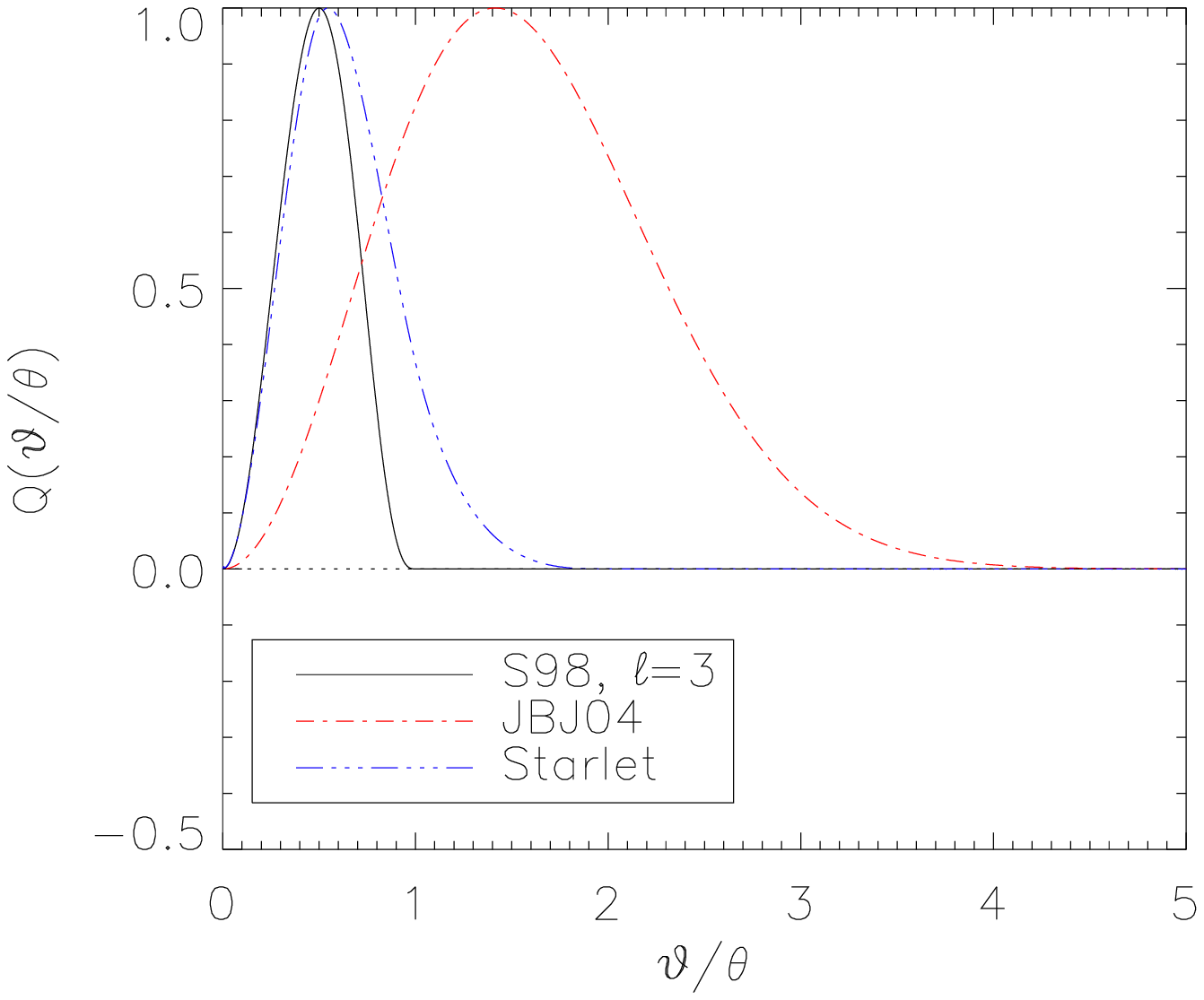}
\caption{Comparison of the $U(\vartheta/\theta)$ (left) and $Q(\vartheta/\theta)$ (right) filter functions commonly used in weak lensing studies and corresponding starlet wavelet functions. Plots are normalised so that each curve attains a peak amplitude of 1. The filters considered are those of Schneider et al. (1998) (taking $\ell = 3$, labelled S98), and Jarvis et al. (2004) (equation \eqref{eq:mexhat} with $b=1/2$, labelled JBJ04). A dotted vertical line in each plot shows the location of the aperture radius, indicating where the filters would be truncated in application of the aperture mass technique. \label{fg:filtersreal}}}
\end{figure}

In Figure \ref{fg:filtersreal}, we compare the aperture mass filter functions defined above with the corresponding starlet wavelet functions in real space. The label \textbf{S98} corresponds to the filters in to equation \eqref{eq:sch} taking $\ell = 3$, and \textbf{JBJ04} corresponds to equation \eqref{eq:mexhat} with $b=1/2$. All filters are renormalised to have a peak value of 1, for comparison purposes. We have not included the filters defined by \cite{vanWaerbeke98} in this comparison, as they are functionally identical to those of \cite{Jarvisetal04} with a different scaling. In practice, they are very similar in scale and shape to the starlet wavelet filters. 

While the S98 $U(\vartheta)$ filter is compensated within an aperture of radius $\theta$, thus yielding an obvious truncation point for this filter function when making maps, neither the starlet or JBJ04 filters show such behaviour. The starlet filter $U(\vartheta)$ filter is, however, compensated within an aperture of radius $2\theta$, and application of the starlet transform algorithm considers the full filter function out to this radius. In contrast, the JBJ04 $Q(\vartheta)$ function tends to zero as $\vartheta\rightarrow\infty$, and the $U(\vartheta)$ filter function is only strictly compensated at this radius.

In practice, a truncation at large $\vartheta$, e.g. $\vartheta_{cut} \sim 5\theta$ would yield minimal error on the solution. However, the larger the truncation radius, the longer the computation time when computing an aperture mass map. Moreover, contamination by edge effects and missing data becomes more problematic as the aperture radius is increased. Two-point and higher order statistics, however, remain unaffected by this truncation, as the effect of the aperture mass filter can be computed analytically in both real and Fourier space for these statistics.

Moreover, statistics such as peak counting are unlikely to be affected by such a truncation. For a non-compensated filter function, the amplitude of structures seen in the resulting $\map$ realisation may be offset by an additive constant proportional to $\int_0^{\vartheta_{cut}}\vartheta U(\vartheta) d\vartheta$ \citep[see][]{S96}. Peak counting is generally performed on the signal to noise of the $M_{ap}$, and as the noise map is generated in the same manner as the signal, it stands to reason that this will not be affected significantly. A similar argument applies to statistical moments of the $\map$ distribution such as the variance, skewness and kurtosis computations performed on the aperture mass map directly, which subtract the mean from the map inherently. 

\subsection{Properties in Fourier Space}
\label{subsec:fourier}

Perhaps more important than their real-space behaviour is the behaviour of these filter functions in Fourier space. The key feature of aperture mass methods is to isolate structures on a particular scale determined by the aperture radius. If this radius is chosen to be much larger than the scale at which the noise is dominant, one may suppress the noise and therefore resolve real structures above the noise. In this way, the aperture mass filters (or, equivalently, the wavelet transform) act as band-pass filters, restricting a given aperture mass map or wavelet scale to a limited range of frequencies in the Fourier domain.

\begin{figure}
\includegraphics[width=0.5\textwidth]{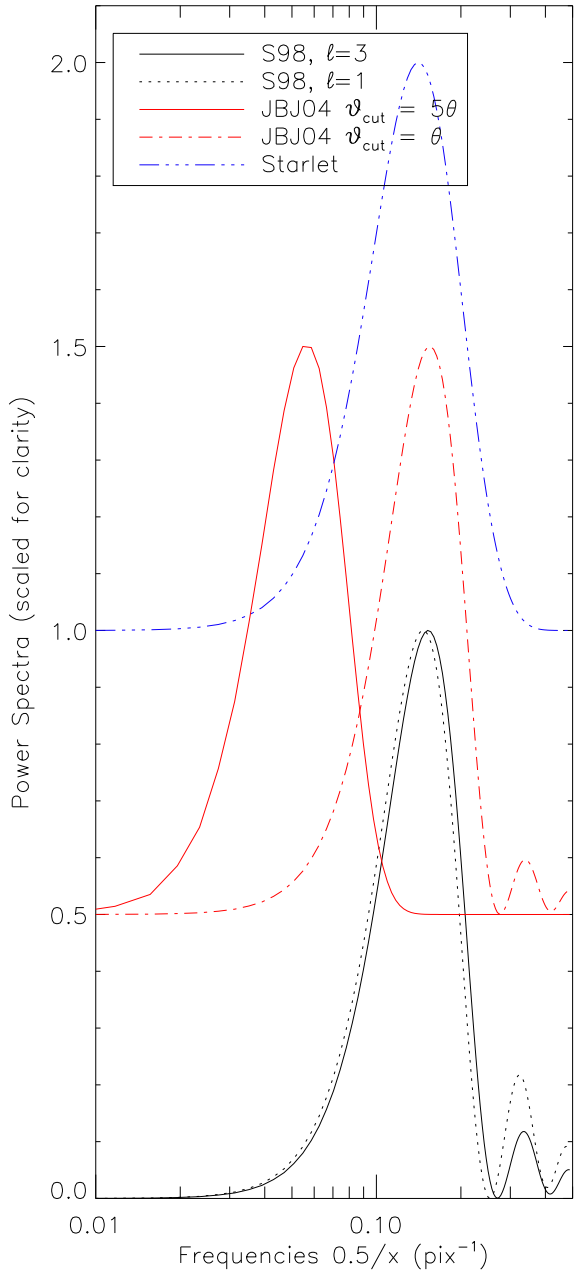}
\caption{Power spectra associated with the aperture mass filters S98, JBJ04 and the starlet wavelet filter. We have plotted the S98 for both $\ell = 3$ as before (solid curve) and $\ell = 1$ (dotted curve) to show the dependence of the power spectrum on the polynomial order $\ell$. For the JBJ04 filter, a truncation has been applied at $\vartheta_{cut} = 5\theta$ (solid curve) and $\vartheta_{cut} = \theta$ (dot-dashed curve). The power spectra for the different filter functions have been offset at intervals of 0.5 along the vertical axis, for clarity.\label{fg:pspec}}
\end{figure}

To assess the response of our aperture mass filters and starlet wavelet transform in Fourier space, we consider the power spectrum of the $U(\vartheta)$ filter function. Ideally, we would like the chosen filter to be localised in the Fourier domain, with a single peak corresponding to the characteristic frequency for the scale being considered, and with no oscillations at high frequencies, as these would result in high-frequency (noisy) information being included in a lower-frequency filtered map.

In order to accurately characterise the behaviour of the filters, including any truncation that is applied, we generate an artificial shear data generated from a null convergence map with a single central delta function, and with no noise in the maps. The aperture mass algorithm is then applied to the resulting shear maps for $\theta = 4$, and the power spectrum of the resulting map computed. For the JBJ04 filter, we consider two truncation radii, at $\vartheta_{cut} = \theta$ and $\vartheta_{cut} = 5\theta$, for illustrative purposes. The wavelet transform is also computed for the map, and the power spectrum for the second wavelet scale computed for comparison. The power spectra are then renormalised to peak at 1. Figure \ref{fg:pspec} shows the resulting power spectra. In the figure, each curve has been offset from the one below by 0.5 along the vertical axis, for clarity.

Broadly, all the filter functions show the expected behaviour. All but one are peaked at approximately the same frequency, indicative of the fact that they are probing structures at a similar scale, and they are all fairly well localised in the frequency domain. The frequency offset seen in the peak of the JBJ04 filter truncated at $\vartheta = 5\theta$ is indicative of the change in the effective width of the filter function as the truncation radius is changed. Oscillations are seen with the JBJ04 filter truncated at $\vartheta_{cut} = \theta$ and the S98 filter at high frequency, but not in the case of the starlet filter or the JBJ04 filter truncated at $\vartheta_{cut} = 5\theta$. 

The oscillations in the strongly truncated JBJ04 case are a result of the fact that the filter function is significantly non-zero and non-compensated within the truncation radius of $\vartheta_{cut} = \theta$. This results in oscillations at high frequency and, as a consequence, a contamination of the resulting $\map$ map with high frequency modes not pertaining to the scale of interest. Clearly, truncations at a large radius, such as $\vartheta_{cut} = 5\theta$ yield negligible oscillations.

Similar oscillations are seen in the S98 filter. This is a direct consequence of the fact that the filter used is a locally-defined function: it is defined to follow the given form only for $\vartheta<\theta$ and artificially set to zero outside this radius by the Heaviside step function. This step function introduces a discontinuity in the filter function, which manifests as oscillations in the Fourier space response of the filter. These oscillations are reduced in amplitude as $\ell$ is increased. 

The starlet filter is very well-behaved at high frequencies, showing no oscillations at any frequency, which results from the fact that the starlet transform algorithm applies no truncation to the starlet filter functions. Therefore, the starlet transform at each scale represents a strictly limited range of frequencies, with no contamination from higher-frequency modes. 

\begin{figure}
\includegraphics[width=0.5\textwidth]{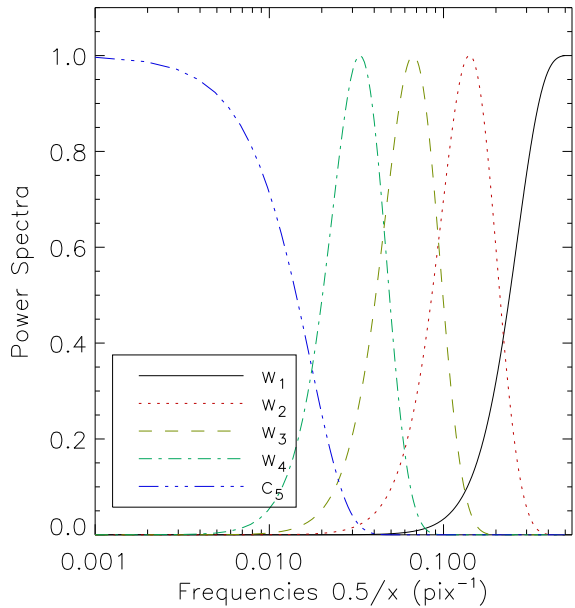}
\caption{Fourier-space response of the wavelet filter as a function of filter scale. Note that at at intermediate scales, the wavelet filter responds as a limited band-pass filter, while at large and small scales, the wavelet transform yields a low- and high-pass filter, respectively. \label{fg:filterstot}}
\end{figure}

The starlet filter offers an additional advantage over aperture mass filtering: the aperture mass filter is a bandpass filter. Because the $\map$ reconstruction is only computed for a discrete set of aperture scales, information on intermediate scales may be lost, particularly at both the small and large frequency extremes. In contrast, the wavelet transform retains information on all scales. The first wavelet scale is effectively a high-pass filter, retaining all the high-frequency information in the image, while the remaining wavelet scales are bandpass filtered as with the $\map$-filtered images. Finally, the wavelet transform retains $c_J$, which encodes the large-scale information in the image, and therefore consists of a smoothed version of the image (or an image with a low-pass filter applied). Figure \ref{fg:filterstot} demonstrates this point for a wavelet transform with $J=5$, showing the Fourier-space response of the wavelet filter as a function of scale $j$.

\subsection{Practical considerations in application}

The starlet transform would seem to offer a clear advantage over traditional $\map$ filtering techniques when it comes to map making, as it requires no truncation of the filter functions and retains information on all scales within the original image. However, in application there are several difficulties that arise, and which must be addressed.

The starlet transform is computed on the convergence map, rather than on the shear catalogue itself. The shear and convergence are related by a straightforward 2D convolution which, in principle, may be inverted directly in Fourier space. However, in practice, this gives rise to significant errors in the convergence estimation, particularly at the edges of the field, arising from the fact that we are considering a finite field sampled over a discrete range of frequencies, while the convolution acts over all space.

This can give rise to $\sim 10\%$ leakage of power into lensing B-modes in the convergence map, which biases the resulting convergence map. Various techniques have been developed to mitigate such effects; most recently, \cite{Deriaz12} have presented a method using a Wavelet-Helmholtz decomposition to separate E- and B-modes in the lensing signal, and to recover the convergence to excellent precision, offering a factor of $\sim 30\times$ reduction in B-mode contamination compared to a direct FFT inversion and $\sim 5\times$ reduction in B-mode contamination compared to the method of \cite{SS96,SS01}, and yielding an RMS error at the percent level in reconstructions. Note also that the \cite{Deriaz12} method may be applied directly to discrete data; i.e. the technique can be applied directly to a shear catalogue, rather than pixelated shear maps.

A further difficulty that arises when attempting to reconstruct the convergence, and compute statistics of the resulting map, is due to missing data in the image. This usually arises due to masking out of bright stars and other contaminating features in the image (such as bad pixels). This lost information is not recovered in the reconstruction, and this biases the statistics computed on the resulting mass map. In \cite{Pires09b}, the authors describe a method by which one can recover information in regions covered by the mask. This method is based on an assumption of sparsity, and the authors find that they are able to recover the power spectrum to a relative error of $0.5-1\%$ using Subaru and CFHT masks, respectively on a $4$ deg$^2$ image, and can recover the equilateral bispectrum to a relative error of $1-3\%$ for the same masks.

Using the techniques described above, it is therefore possible to compute maps of the convergence to sufficient accuracy to be able to compute associated statistics on the resulting maps, and thereby place constraints on cosmological parameters in a similar manner to methods applied directly to the shear catalogue. 

\section{Algorithm Speeds}
\label{sec:timing}
\subsection{Map making}

Generating an aperture mass map generally involves brute force application of equation \eqref{eq:shearmap} to a given dataset. Such an algorithm has complexity $\propto \mathcal{O}(N^2\vartheta_{cut}^2)$, where $N \times N$ is the dimension of the image, and $\vartheta_{cut}$ is the chosen truncation radius. This scaling means that for large apertures or, equivalently, for high-resolution images, the algorithm may prove to be very time-consuming. Estimation of the noise in aperture mass studies is generally carried out by randomising the data and repeating the measurement many times. Therefore it is of great importance to be able to compute the aperture mass for a given image rapidly.

The starlet wavelet transform algorithm is of complexity $\propto \mathcal{O}(N^2J)$, where $J$ is the number of scales considered, and is limited by $N \ge 2^{J}$. This means that the processing time for the wavelet transform algorithm is sensitive only to the number of scales considered, rather than the size of the filter functions involved, and depends linearly on this number. 

\begin{figure}
\includegraphics[width=0.5\textwidth]{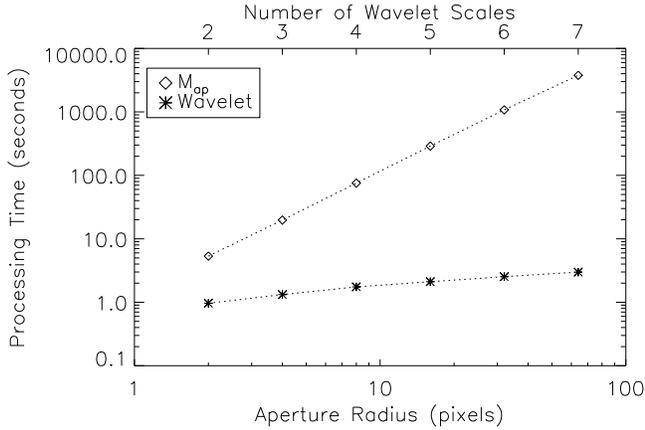}
\caption{Comparison of the processing time of the aperture mass algorithm and the starlet transform algorithm to analyse an image of $1024 \times 1024$ pixels on a $2\times2.66$GHz Intel Xeon Dual-Core processor. The top axis represents the total number $J$ of $1024 \times 1024$ pixel arrays computed by the wavelet transform algorithm, including the smooth continuum map.\label{fg:timing}}
\end{figure}

In Figure \ref{fg:timing}, we compare the processing time for the aperture mass algorithm and the starlet transform algorithm\footnote{available from http://irfu.cea.fr/Ast/mrlens\_software.php}, both programmed in C++, to analyse an image of $1024 \times 1024$ pixels on a $2\times2.66$GHz Intel Xeon Dual-Core processor. We consider aperture scales $\theta = [2, 4, 8, 16, 32, 64]$, which correspond to $J = j_{\max}+1 = [2, 3, 4, 5, 6, 7]$ wavelet scales in the wavelet transform. In the aperture mass algorithm, the filters are truncated at a radius of $\vartheta_{cut} = \theta$. For the application of filters such as the JBJ04 filter, which necessitates truncation at a much larger radius, we expect the computation time to be roughly an order of magnitude longer. Even at the smallest aperture radius, the wavelet transform is $\sim 5\times$ faster than the aperture mass algorithm. At $\theta = 64$ pixels, the wavelet transform is $\sim 1200 \times$ faster than the aperture mass algorithm. Note that the wavelet transform for a given value of $J$ simultaneously computes the wavelet transform at all scales $2^j,\ 0<j\le J-1$, in addition to the smoothed continuum map $c_{J}$, whilst the aperture mass algorithm computes the transform at a single scale $\theta$. We note further that the computational time for the wavelet transform for $J = 7$ wavelet scales is still a factor of $\sim 2\times$ less than the computational time for the aperture mass algorithm at $\theta = 2$ pixels.

The starlet transform algorithm therefore offers a clear and significant time advantage over the aperture mass algorithm for all scales of interest. Note that wavelets do not restrict our data analysis only to dyadic scales. Indeed some other wavelet transform algorithms, such as that based on fractional splines  \citep{unser2000}, allow us to use any intermediate wavelet scale.

\subsection{Computing higher-order $\map$ statistics}

When computing higher-order statistics such as the variance, skewness or kurtosis of the aperture mass statistic, one can choose to work directly on the shear field, and compute the statistics via the $n$-point correlation function. Computation of such statistics using a naive approach usually results in an algorithm with complexity $\mathcal{O}(N_{gal}^n)$, where $N_{gal}$ is the number of galaxies being considered and $n$ is the order of the correlation function. On large fields, such a computation can be prohibitive.

In recent years, tree codes have been employed to speed up computation of $n$-point correlation functions. Typical tree codes to compute $n$-point correlation functions are $\mathcal{O}(N_{gal}\log(N_{gal}))$ \citep{Zhang05} on a shear catalogue. For a single Euclid exposure of $0.5$ deg$^2$, we can expect $N_{gal} \sim 54,000$ ($30$ galaxies/arcmin$^2$) - $180,000$ ($100$ galaxies/arcmin$^2$). Tree codes exist that act on pixelated data \citep[e.g.][]{Eriksen04} which run at $\mathcal{O}(N_{pix}^2n_{bin})$ where $N_{pix}^2$ is the total number of pixels and $n_{bin}$ is the number of bins in the correlation function. For a Euclid exposure, assuming pixels of 1 arcminute, we have $N_{pix}^2=1800$, and $n_{bin}$ will be dependent on the required resolution of the correlation function. Typically, statistics will be computed on much larger fields, however (for example, \cite{Pires09a} compute statistics on a field of $3.95 \times 3.95$ deg$^2$, which is a scale-up of $\sim 30$ in area).

The wavelet method acts on pixellated data, and is $\mathcal{O}(N^2J)$ in computation time, so our algorithm will be comparable for computation of 2-point statistics, if the 2-point correlation function is computed on pixellated data, but if a shear catalogue is used, we will have a faster algorithm by at least an order of magnitude. For higher-order statistics, this advantage is even more pronounced. Furthermore, while optimised software is freely and publicly available to compute the wavelet transform, such optimised software is not available for $n$-point correlation functions.

Note that \cite{Jarvisetal04} show an $\mathcal{O}(N_{gal})$ complexity for the computation of n-point statistics with their algorithm, with no $\log(N_{gal})$ factor. However, for $n \ge 3$, the prefactor becomes quite large, and for $n=3$ their algorithm was unable to reach the theoretical $\mathcal{O}(N_{gal}\log(N_{gal}))$ scaling at $N_{gal}=10^6$. This means that their algorithm may be competitive for computation of two-point statistics, but for higher order statistics, methods acting on pixellated data -- such as the wavelet method described above -- would be advantageous.

\subsection{Further advantages of wavelets}

In \cite{SPR06}, the authors describe a method for reconstructing the convergence from shear measurements using wavelets. Using their method, it is possible to carry out an explicit denoising of the convergence field using thresholding based on a False Discovery Rate method. This allows them to derive robust detection levels in wavelet space, and to produce high-fidelity denoised mass maps. In \cite{Pires09a,pires11}, it is shown that more non-gaussian information can be extracted from these wavelet-denoised maps as compared to the shear, convergence or $\map$-filtered maps, and therefore that tighter constraints may be placed on the cosmological model using the wavelet method.

In addition, wavelet-based methods offer more flexibility than aperture mass filters. Whilst we have presented here only the starlet wavelet function, many other wavelet dictionaries may be used. The starlet filter seems ideal for lensing studies, due to its similarity to the JBJ04 filter presented here, which was deemed to be optimal in \cite{Zhang03}. However, different dictionaries may be optimal in different applications; for example, if one were attempting to study filamentary structure, ridgelets or curvelets might be a more appropriate basis. The vastness of the wavelet libraries and the public availability of fast algorithms to compute these transforms are major strengths of wavelet-based approaches.

\section{Summary and discussion}
\label{sec:summary}

In this paper, we have compared the aperture mass statistic, a commonly-used tool in weak lensing studies, with the 2D wavelet transform using a starlet wavelet. We have demonstrated that the aperture mass technique is formally identical to the wavelet transform evaluated at a specific scale. 

We have considered two common forms for the aperture mass filter function, and compared it to the equivalent starlet functions. One of these aperture mass filters -- labelled JBJ04 -- is considered to be the optimal aperture mass filter for skewness studies \citep{Zhang03}, and is thus considered to be the state-of-the-art in this field. We outline three criteria for the aperture mass filters: 
\begin{enumerate}
\item They should be local in real space.
\item They should be compensated within the aperture radius being considered.
\item They should be local and non-oscillatory in Fourier space. 
\end{enumerate}
We demonstrate that none of the commonly used aperture mass filter functions meet all three criteria simultaneously, whilst the starlet -- by definition -- meets all of the criteria.

Lastly, we show that the starlet transform algorithm, which forms part of a publicly available software package containing numerous tools for weak lensing, dramatically outperforms a brute force application of the aperture mass technique, with speed-up factors of $5-\sim1200$ seen for the processing of a single aperture mass reconstruction, depending on the scale considered.

Given that the aperture mass technique is identical to the wavelet transform, that the starlet wavelet function is more appropriately suited to aperture mass studies, and that the starlet transform algorithm offers such a great advantage in terms of processing time over the standard aperture mass algorithm, we argue that the wavelet transform should be the preferred method in future weak lensing studies involving use of the aperture mass method.

A further advantage of this wavelet method is that fast and effective software is available to carry out explicit denoising in the wavelet domain \citep{SPR06}. In \cite{Pires09a, pires11}, it has been shown that further cosmological information can be extracted from noisy convergence maps if such denoising is applied to the convergence maps, as the starlet wavelet transform is particularly sensitive to non-gaussian structures. In addition, a multitude of wavelet libraries are available, and may be chosen to be optimal for a given problem. Thus, wavelet methods have an inherent flexibility in application.

\section{acknowledgments}
The authors would like to thank Martin Kilbinger for very useful discussions, and the anonymous referee, for helpful comments on this work.
This work is supported by the European Research Council grant SparseAstro (ERC-228261). 

\bibliography{refs}

\begin{thebibliography}{}

\bibitem[\protect\citeauthoryear{{Crittenden}, {Natarajan}, {Pen} \&
  {Theuns}}{{Crittenden} et~al.}{2002}]{Crittenden02}
{Crittenden} R.~G.,  {Natarajan} P.,  {Pen} U.-L.,    {Theuns} T.,  2002, ApJ,
  568, 20

\bibitem[\protect\citeauthoryear{{Curto}, {Mart{\'{\i}}nez-Gonz{\'a}lez} \&
  {Barreiro}}{{Curto} et~al.}{2011}]{martinez11}
{Curto} A.,  {Mart{\'{\i}}nez-Gonz{\'a}lez} E.,    {Barreiro} R.~B.,  2011,
  MNRAS, 412, 1038

\bibitem[\protect\citeauthoryear{{Deriaz}, {Starck} \& {Pires}}{{Deriaz}
  et~al.}{2012}]{Deriaz12}
{Deriaz} E.,  {Starck} J.-L.,    {Pires} S.,  2012, ArXiv e-prints

\bibitem[\protect\citeauthoryear{{Dietrich} \& {Hartlap}}{{Dietrich} \&
  {Hartlap}}{2010}]{DH10}
{Dietrich} J.~P.,  {Hartlap} J.,  2010, MNRAS, 402, 1049

\bibitem[\protect\citeauthoryear{Eriksen, Lilje, Banday \& G—rski}{Eriksen
  et~al.}{2004}]{Eriksen04}
Eriksen H.~K.,  Lilje P.~B.,  Banday A.~J.,    G—rski K.~M.,  2004, The
  Astrophysical Journal Supplement Series, 151, 1

\bibitem[\protect\citeauthoryear{{Jarvis}, {Bernstein} \& {Jain}}{{Jarvis}
  et~al.}{2004}]{Jarvisetal04}
{Jarvis} M.,  {Bernstein} G.,    {Jain} B.,  2004, MNRAS, 352, 338

\bibitem[\protect\citeauthoryear{Jin, Starck, Donoho, Aghanim \& Forni}{Jin
  et~al.}{2005}]{starck:jin05}
Jin J.,  Starck J.-L.,  Donoho D.,  Aghanim N.,    Forni O.,  2005, Eurasip
  Journal, 15, 2470

\bibitem[\protect\citeauthoryear{{Kilbinger} \& {Schneider}}{{Kilbinger} \&
  {Schneider}}{2005}]{KS05}
{Kilbinger} M.,  {Schneider} P.,  2005, A\&A, 442, 69

\bibitem[\protect\citeauthoryear{{McEwen}, {Hobson}, {Lasenby} \&
  {Mortlock}}{{McEwen} et~al.}{2008}]{McEwen08a}
{McEwen} J.~D.,  {Hobson} M.~P.,  {Lasenby} A.~N.,    {Mortlock} D.~J.,  2008,
  MNRAS, 388, 659

\bibitem[\protect\citeauthoryear{{Pires}, {Leonard} \& {Starck}}{{Pires}
  et~al.}{2012}]{pires11}
{Pires} S.,  {Leonard} A.,    {Starck} J.-L.,  2012, MNRAS submitted

\bibitem[\protect\citeauthoryear{{Pires}, {Starck}, {Amara},
  {R{\'e}fr{\'e}gier} \& {Teyssier}}{{Pires} et~al.}{2009}]{Pires09a}
{Pires} S.,  {Starck} J.-L.,  {Amara} A.,  {R{\'e}fr{\'e}gier} A.,
  {Teyssier} R.,  2009, A\&A, 505, 969

\bibitem[\protect\citeauthoryear{{Pires}, {Starck}, {Amara}, {Teyssier},
  {R{\'e}fr{\'e}gier} \& {Fadili}}{{Pires} et~al.}{2009}]{Pires09b}
{Pires} S.,  {Starck} J.-L.,  {Amara} A.,  {Teyssier} R.,  {R{\'e}fr{\'e}gier}
  A.,    {Fadili} J.,  2009, MNRAS, 395, 1265

\bibitem[\protect\citeauthoryear{{Schneider}}{{Schneider}}{1996}]{S96}
{Schneider} P.,  1996, Monthly Notices of the Royal Astronomical Society, 283,
  837

\bibitem[\protect\citeauthoryear{{Schneider}, {van Waerbeke}, {Jain} \&
  {Kruse}}{{Schneider} et~al.}{1998}]{Schneideretal98}
{Schneider} P.,  {van Waerbeke} L.,  {Jain} B.,    {Kruse} G.,  1998, Monthly
  Notices of the Royal Astronomical Society, 296, 873

\bibitem[\protect\citeauthoryear{{Schneider}, {van Waerbeke}, {Kilbinger} \&
  {Mellier}}{{Schneider} et~al.}{2002}]{Schneideretal02}
{Schneider} P.,  {van Waerbeke} L.,  {Kilbinger} M.,    {Mellier} Y.,  2002,
  A\&A, 396, 1

\bibitem[\protect\citeauthoryear{{Seitz} \& {Schneider}}{{Seitz} \&
  {Schneider}}{1996}]{SS96}
{Seitz} S.,  {Schneider} P.,  1996, Astronomy and Astrophysics, 305, 383

\bibitem[\protect\citeauthoryear{{Seitz} \& {Schneider}}{{Seitz} \&
  {Schneider}}{2001}]{SS01}
{Seitz} S.,  {Schneider} P.,  2001, Astronomy and Astrophysics, 374, 740

\bibitem[\protect\citeauthoryear{{Starck}, {Murtagh} \& {Bijaoui}}{{Starck}
  et~al.}{1998}]{smb98}
{Starck} J.,  {Murtagh} F.~D.,    {Bijaoui} A.,  1998, {Image Processing and
  Data Analysis}.
Cambridge University Press

\bibitem[\protect\citeauthoryear{Starck, Aghanim \& Forni}{Starck
  et~al.}{2004}]{starck:sta03_1}
Starck J.-L.,  Aghanim N.,    Forni O.,  2004, A\&A, 416, 9

\bibitem[\protect\citeauthoryear{{Starck} \& {Murtagh}}{{Starck} \&
  {Murtagh}}{2006}]{starck:book06}
{Starck} J.-L.,  {Murtagh} F.,  2006, {Astronomical Image and Data Analysis}.
Springer

\bibitem[\protect\citeauthoryear{Starck, Murtagh \& Bertero}{Starck
  et~al.}{2011}]{starck:book11}
Starck J.-L.,  Murtagh F.,    Bertero M.,  2011, in Handbook of Mathematical
  Methods in Imaging.
Springer

\bibitem[\protect\citeauthoryear{Starck, Murtagh \& Fadili}{Starck
  et~al.}{2010}]{starck:book10}
Starck J.-L.,  Murtagh F.,    Fadili M.,  2010, Sparse Image and Signal
  Processing.
Cambridge University Press

\bibitem[\protect\citeauthoryear{{Starck}, {Pires} \&
  {R{\'e}fr{\'e}gier}}{{Starck} et~al.}{2006}]{SPR06}
{Starck} J.-L.,  {Pires} S.,    {R{\'e}fr{\'e}gier} A.,  2006, A\&A, 451, 1139

\bibitem[\protect\citeauthoryear{Unser \& Blu}{Unser \& Blu}{2000}]{unser2000}
Unser M.,  Blu T.,  2000, SIAM Review, 42, 43

\bibitem[\protect\citeauthoryear{{van Waerbeke}}{{van
  Waerbeke}}{1998}]{vanWaerbeke98}
{van Waerbeke} L.,  1998, A\&A, 334, 1

\bibitem[\protect\citeauthoryear{{Vielva}, {Mart{\'{\i}}nez-Gonz{\'a}lez},
  {Barreiro}, {Sanz} \& {Cay{\'o}n}}{{Vielva} et~al.}{2004}]{gauss:vielva04}
{Vielva} P.,  {Mart{\'{\i}}nez-Gonz{\'a}lez} E.,  {Barreiro} R.~B.,  {Sanz}
  J.~L.,    {Cay{\'o}n} L.,  2004, Astrophysical Journal, 609, 22

\bibitem[\protect\citeauthoryear{{Vielva}, {Wiaux},
  {Mart{\'{\i}}nez-Gonz{\'a}lez} \& {Vandergheynst}}{{Vielva}
  et~al.}{2006}]{vielva06}
{Vielva} P.,  {Wiaux} Y.,  {Mart{\'{\i}}nez-Gonz{\'a}lez} E.,
  {Vandergheynst} P.,  2006, New Astronomy Review, 50, 880

\bibitem[\protect\citeauthoryear{Zhang \& Pen}{Zhang \& Pen}{2005}]{Zhang05}
Zhang L.~L.,  Pen U.-L.,  2005, New Astronomy, 10, 569

\bibitem[\protect\citeauthoryear{{Zhang}, {Pen}, {Zhang} \& {Dubinski}}{{Zhang}
  et~al.}{2003}]{Zhang03}
{Zhang} T.-J.,  {Pen} U.-L.,  {Zhang} P.,    {Dubinski} J.,  2003, ApJ, 598,
  818

\end{thebibliography}

\label{lastpage}

\end{document}